\documentclass[11pt]{article}
\usepackage[a4paper,margin=1in]{geometry}
\usepackage[T1]{fontenc}
\usepackage[utf8]{inputenc}
\usepackage{lmodern}
\usepackage{amsmath,amssymb,amsthm,mathtools}
\usepackage{microtype}
\usepackage{enumitem}
\usepackage{hyperref}
\usepackage[nameinlink,noabbrev]{cleveref}
\usepackage{tikz}
\usetikzlibrary{arrows.meta,calc,decorations.pathmorphing,positioning}

\hypersetup{colorlinks=true,linkcolor=blue,citecolor=blue,urlcolor=blue}

\newtheorem{theorem}{Theorem}[section]
\newtheorem{proposition}[theorem]{Proposition}
\newtheorem{lemma}[theorem]{Lemma}
\newtheorem{corollary}[theorem]{Corollary}
\theoremstyle{definition}
\newtheorem{definition}[theorem]{Definition}
\newtheorem{remark}[theorem]{Remark}

\newcommand{\R}{\mathbb{R}}
\newcommand{\Z}{\mathbb{Z}}
\newcommand{\Mink}{\mathbb{M}^{1,1}}
\newcommand{\etaM}{\eta}
\newcommand{\Ireg}{\mathrm{I}}
\newcommand{\IIreg}{\mathrm{II}}
\newcommand{\IIIreg}{\mathrm{III}}
\newcommand{\IVreg}{\mathrm{IV}}

\newcommand{\cX}{\mathcal{X}}
\newcommand{\cM}{\mathcal{M}}

\newcommand{\into}{\hookrightarrow}
\newcommand{\angles}[1]{\langle #1\rangle}

\title{Coverings and Non-Hausdorff Extensions of Misner Spacetime}
\author{}
\date{}

\begin{document}
\begin{center}
{\LARGE\bfseries Coverings and Non-Hausdorff Extensions of Misner Spacetime\par}
\vspace{1em}
{\large N. E. Rieger\par}
\vspace{0.5em}
{\small
Mathematics Department, University of Southern California, Los Angeles, CA 90089, USA\par
Mathematics Department, University of California, Irvine, Rowland Hall, Irvine, CA 92697, USA\par
\texttt{n.rieger@uci.edu}\par}
\end{center}

\vspace{1em}

~

\noindent
This article is based on research originally conducted as part of a project during 2016--2018 under the supervision of Kip S. Thorne.

\begin{abstract}
Misner spacetime is obtained by quotienting a timelike wedge of two-dimensional Minkowski spacetime by a discrete boost. The familiar Hausdorff extensions and the Hawking--Ellis non-Hausdorff extension are classical, but the passage from covering constructions of the punctured Minkowski plane to genuine extensions of Misner spacetime is subtler than is often stated. In this article we separate systematically the notions of covering and extension, classify the connected coverings of the punctured model that are compatible with the boost action, construct the induced quotient spacetimes, and exhibit explicit embeddings of Misner spacetime into each of them. This yields a natural family consisting of the Hawking--Ellis extension, its universal-cover analogue, and the intermediate finite cyclic coverings. We prove a precise non-Hausdorffness statement for the punctured quotient, formulate and prove a classification theorem for the resulting family within the covering-compatible class, and identify a causal adjacency invariant distinguishing the finite-sheeted and universal-cover cases. Finally, we compare these spacetimes with two-dimensional Schwarzschild-type metrics from the viewpoint of isocausality. 
\end{abstract}


\section{Introduction}

\begin{figure}
\centering
\includegraphics[width=0.5\columnwidth]{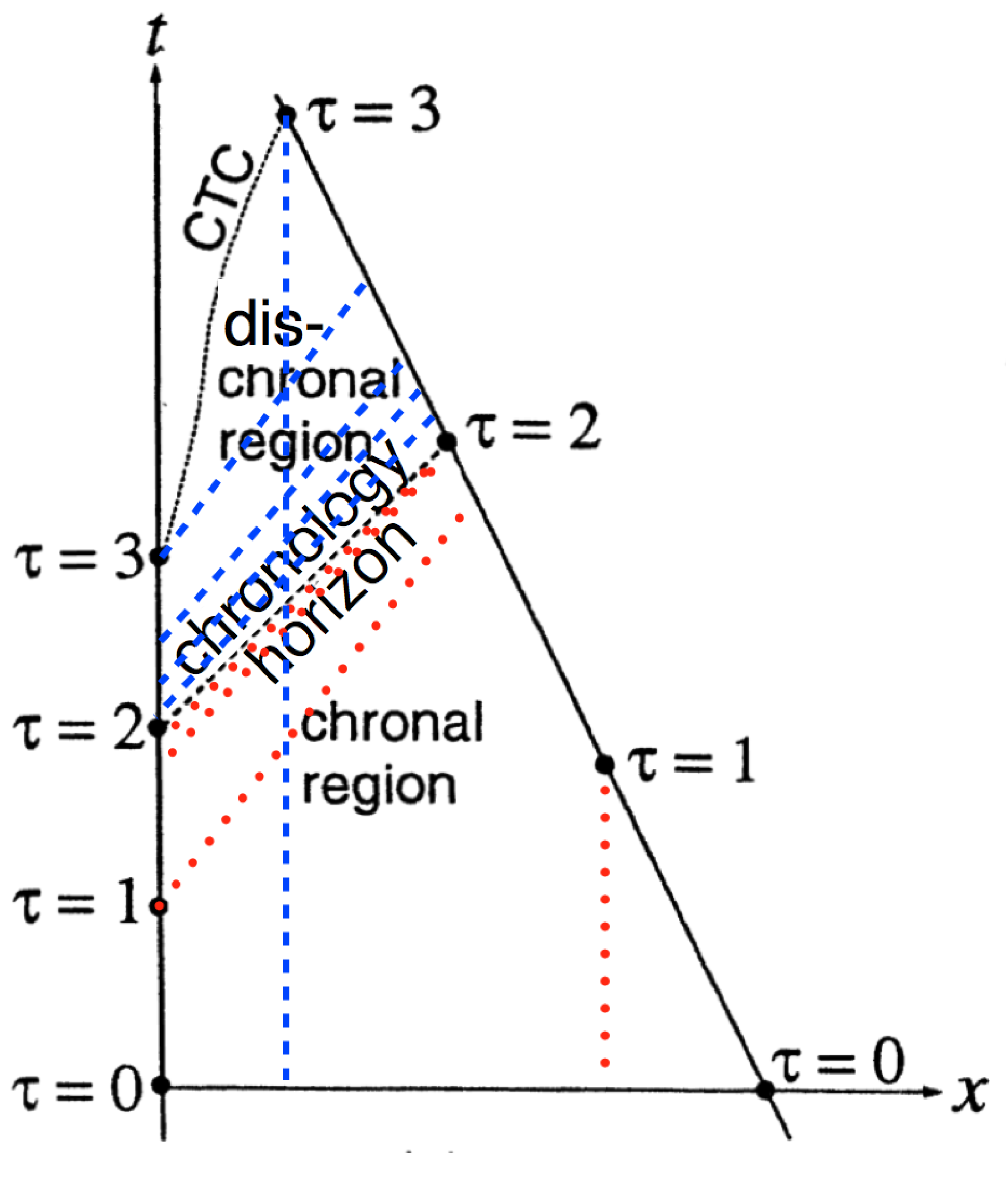}
\caption{Misner spacetime depicted as a ``room'' in flat two-dimensional spacetime with identified walls moving toward each other. The slanted wall is identified with the vertical wall by the boost. The chronology horizon separates the lower chronal region from the upper dischronal region containing closed timelike curves.}
\label{fig:MovingWall}
\end{figure}

Misner spacetime remains a basic model for chronology horizons, incomplete geodesics, and the tension between local flatness and global causal pathologies~\cite{mi}. Because the geometry is fully explicit, one can test broad ideas---extension theory, quotient constructions, causal comparison, and the role of non-Hausdorff manifolds---in a setting where all constructions can be written down concretely; see, for example,~\cite{he,hk,th,tho,la,su,lo,mv}. A standard ``moving walls'' picture is shown in \Cref{fig:MovingWall}. 

The classical description is well known. Starting from a timelike wedge in $\Mink$ and quotienting by a discrete boost, one obtains a smooth Hausdorff Lorentzian cylinder whose chronology horizon is incomplete. Enlarging across one component of the future chronology horizon yields one of the two inequivalent Hausdorff extensions discussed by Hawking and Ellis~\cite{he}; enlarging across both sides simultaneously yields the punctured Minkowski plane modulo the same boost action, producing the standard non-Hausdorff extension.

The purpose of the present paper is to isolate the natural family of further non-Hausdorff extensions generated by connected coverings of the punctured Minkowski plane, and to describe that family with precise hypotheses and explicit embeddings. The point is not merely that the punctured plane has many coverings. Rather, each connected covering supports a canonical lifted boost action, and the associated quotient gives a Lorentzian manifold into which Misner spacetime embeds isometrically as an open submanifold. In this way one obtains a family of genuine extensions, not just a family of covering spaces.

The main results are as follows.
\begin{enumerate}[label=(\roman*)]
    \item We separate cleanly the three levels of structure: the wedge model defining Misner spacetime, the punctured Minkowski plane, and its connected coverings.
    \item We classify all connected coverings of the punctured model that are compatible with the boost action and prove that, up to equivalence, they are indexed by $n\in\mathbb{N}\cup\{\infty\}$.
    \item For each such covering we construct a canonical quotient spacetime $E_n$ and an explicit isometric embedding $\iota_n\colon \cM\into E_n$ onto an open subset.
    \item We prove a causal adjacency theorem: in the universal-cover extension the chronal sectors form a bi-infinite chain, while in the finite-sheeted cases they form a directed $n$-cycle. This graph is a causal invariant.
    \item We analyze isocausality with two-dimensional Schwarzschild-type metrics. Local isocausality holds on chronal simply connected regions, whereas global isocausality fails for any Misner-type extension carrying closed timelike curves when compared with a chronological Schwarzschild region. We also place this beside the known Misner--pseudo-Schwarzschild isocausality established in~\cite{rieger}.
\end{enumerate}

The article is organized as follows. \Cref{sec:basic} fixes notation and states precise definitions. \Cref{sec:classical} reviews the classical Hausdorff and non-Hausdorff extensions and gives a proof of the Hausdorff failure mechanism. \Cref{sec:coverings} classifies the connected coverings of the punctured plane and lifts the boost action. \Cref{sec:extensions} shows that the quotient spacetimes are genuine extensions of Misner spacetime and formulates a classification theorem within the covering-compatible class. \Cref{sec:causal} proves the causal adjacency theorem and derives a causal invariant. \Cref{sec:isocausality} treats the comparison with Schwarzschild-type metrics.

\section{Misner spacetime, boosts, and precise notions}\label{sec:basic}

\subsection{The basic quotient}

Let $(\Mink,\etaM)$ be two-dimensional Minkowski spacetime with standard coordinates $(t,x)$ and null coordinates
\[
 u=t-x,\qquad v=t+x,
\]
so that
\[
 \etaM=-du\,dv.
\]
Fix $\lambda>0$ and define the Lorentz boost
\begin{equation}\label{eq:boost}
 b_\lambda(u,v)=\bigl(e^{\lambda}u,e^{-\lambda}v\bigr).
\end{equation}
Let
\[
 G:=\angles{b_\lambda}\cong \Z
\]
be the infinite cyclic group generated by this single isometry. We reserve $b_\lambda$ for the boost itself and $G$ for the group it generates.

The origin $Q=(0,0)$ is the unique fixed point of the boost action. Removing the two null axes through $Q$ decomposes $\Mink\setminus\{Q\}$ into the four standard wedges
\[
\Ireg=\{u<0,v<0\},\quad \IIreg=\{u<0,v>0\},\quad \IIIreg=\{u>0,v<0\},\quad \IVreg=\{u>0,v>0\}.
\]
We take \emph{Misner spacetime} to be the quotient
\[
\cM:=\Ireg/G.
\]
Because the action of $G$ on $\Ireg$ is free and properly discontinuous, $\cM$ is a smooth Hausdorff Lorentzian cylinder.

In $\Ireg$ it is convenient to introduce adapted coordinates
\begin{equation}\label{eq:Tphi}
 T=-uv>0,\qquad \phi=\log\!\left(-\frac{v}{u}\right)\mod \lambda,
\end{equation}
for which the metric becomes, after the harmless rescaling $\phi\mapsto 2\phi$,
\begin{equation}\label{eq:misnermetric}
 g=-2\,dT\,d\phi-T\,d\phi^2.
\end{equation}
The chronology horizon corresponds to $T=0$.

\subsection{Coverings and extensions}

We now fix the terminology used throughout.

\begin{definition}
A Lorentzian manifold $(\widetilde X,\widetilde g)$ is a \emph{covering spacetime} of $(X,g)$ if there exists a covering map $p\colon \widetilde X\to X$ such that $p^*g=\widetilde g$.
\end{definition}

\begin{definition}
A Lorentzian manifold $(X',g')$ is an \emph{extension} of $(X,g)$ if there exists an isometric embedding
\[
\iota\colon (X,g)\into (X',g')
\]
onto a proper open subset of $X'$.
\end{definition}

A cover need not be an extension, and an extension need not itself be a covering space of anything. In the present problem the punctured Minkowski plane and its further coverings first appear as covering spaces; only after passing to suitable quotients and constructing the embedding of $\cM$ do they become extensions of Misner spacetime.

\section{Classical Hausdorff and non-Hausdorff extensions}\label{sec:classical}

The standard Hausdorff extensions are obtained by enlarging $\Ireg$ across exactly one component of the future chronology horizon. Set
\[
\cM_+:=(\Ireg\cup \IIreg)/G,
\qquad
\cM_-:=(\Ireg\cup \IIIreg)/G.
\]
These are the two inequivalent maximal Hausdorff extensions discussed by Hawking and Ellis~\cite{he}. Their universal covers are simply the half-planes $\Ireg\cup\IIreg$ and $\Ireg\cup\IIIreg$.

\begin{figure}
\centering
\includegraphics[width=0.5\columnwidth]{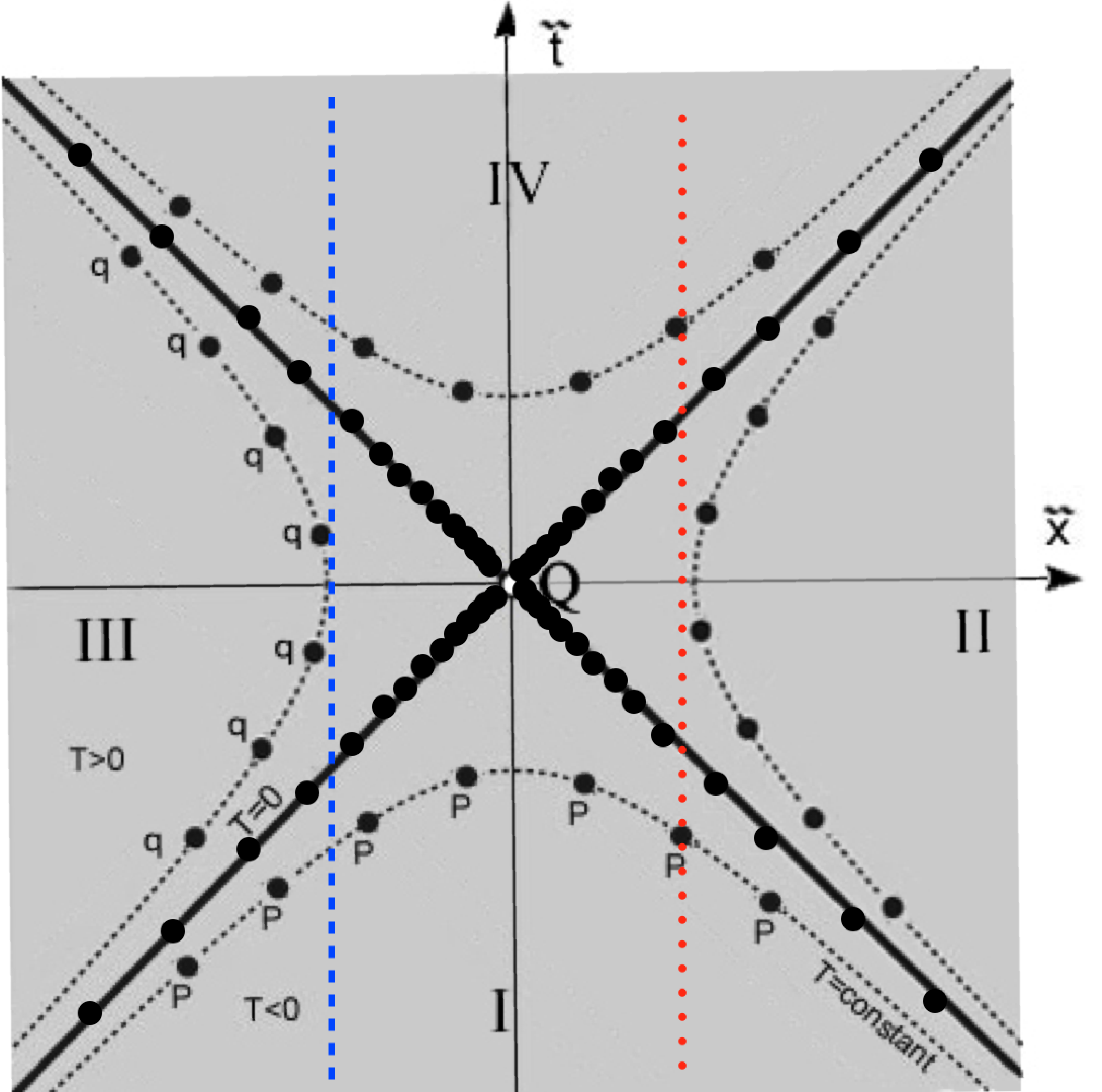}
\caption{Minkowski spacetime with the four standard wedges determined by the null axes through $Q$. Boost-related points lie on the same hyperbola, and the null diagonals are the chronology horizons along which the $G$-orbits accumulate in the non-Hausdorff quotient.}
\label{fig:MinkowskiBoost}
\end{figure}

The simultaneous enlargement across both sides leads to the punctured plane
\[
\cX:=\Mink\setminus\{Q\},
\]
endowed with the induced flat metric. The action of $G$ on $\cX$ remains free but is no longer properly discontinuous, because boost orbits accumulate along the null axes. The quotient
\[
E_1:=\cX/G
\]
is the standard Hawking--Ellis non-Hausdorff extension. It is smooth and $T_1$, but not Hausdorff (\Cref{fig:MinkowskiBoost}).

\begin{proposition}\label{prop:nonhaus}
The quotient $E_1=\cX/G$ is non-Hausdorff. More precisely, let $x_+\in E_1$ and $x_-\in E_1$ be the images of two points lying on the distinct future null half-lines $\{u=0,v>0\}$ and $\{v=0,u>0\}$ with the same boost parameter. Then every pair of saturated neighborhoods of $x_+$ and $x_-$ intersects.
\end{proposition}

\begin{proof}
Let $\pi\colon \cX\to E_1$ be the quotient map, and choose representatives
\[
 p_+=(0,v_0),\qquad p_-=(u_0,0),\qquad u_0,v_0>0,
\]
with matched boost parameter. Let $U_+$ and $U_-$ be arbitrarily small neighborhoods of $p_+$ and $p_-$ in $\cX$.

Under the boost~\eqref{eq:boost}, the image $b_\lambda^k(U_+)$ becomes exponentially narrower in the $u$-direction and exponentially broader in the $v$-direction as $k\to +\infty$. Conversely, $b_\lambda^{-k}(U_-)$ becomes exponentially broader in the $u$-direction and exponentially narrower in the $v$-direction. For sufficiently large $k$, the two sets $b_\lambda^k(U_+)$ and $b_\lambda^{-k}(U_-)$ intersect in the quadrant $\IVreg$. Hence the saturated sets
\[
G\cdot U_+=\bigcup_{m\in\Z}b_\lambda^m(U_+),
\qquad
G\cdot U_-=\bigcup_{m\in\Z}b_\lambda^m(U_-)
\]
intersect. Their images under $\pi$ therefore cannot be disjoint. Since the original neighborhoods were arbitrary, $x_+$ and $x_-$ are not separable.

This is precisely the mechanism behind the failure of Hausdorffness. One does not attempt to separate the origin from the horizon points: the origin is not present in the quotient at all.
\end{proof}

\section{Coverings of the punctured plane and lifted boost actions}\label{sec:coverings}

\subsection{Classification of connected coverings}

Topologically, $\cX=\Mink\setminus\{Q\}$ deformation retracts onto a circle. Hence
\[
\pi_1(\cX)\cong \Z.
\]
By the classification of connected coverings, up to equivalence there is exactly one connected covering corresponding to each subgroup of $\Z$, namely $n\Z$ for $n\in\mathbb{N}$, together with the trivial subgroup $\{0\}$.

We denote by
\[
 p_n\colon S_n\to \cX\qquad (n\in\mathbb{N})
\]
the connected covering associated with $n\Z$, and by
\[
 p_\infty\colon S_\infty\to \cX
\]
the universal covering. Thus $S_1=\cX$, every finite $S_n$ is a cyclic $n$-sheeted covering of $\cX$, and $S_\infty$ is simply connected.

\begin{figure}
\centering
\includegraphics[width=1.05\columnwidth]{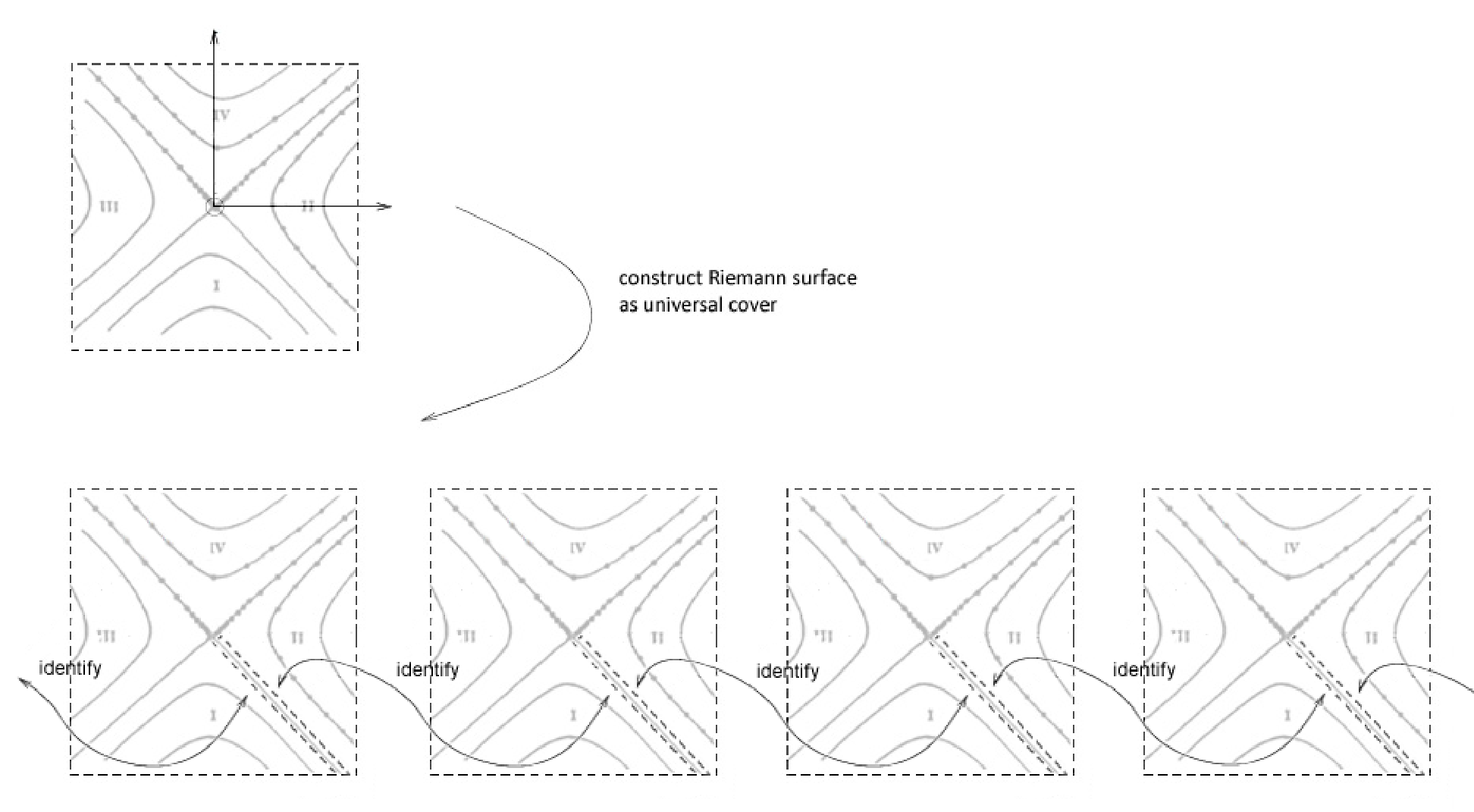}
\caption{Schematic cut-and-glue construction of connected coverings of the punctured Minkowski plane. Successive copies of $\cX$ are cut along a chosen future null half-line and glued sheet-to-sheet. Identifying after $n$ sheets produces the finite cyclic covering $S_n$; leaving the chain open gives the universal cover $S_\infty$.}
\label{fig:cutglue}
\end{figure}

A concrete model is obtained by cutting $\cX$ along one future null half-line and gluing copies successively; see \Cref{fig:cutglue}.\footnote{An alternative description uses split-complex numbers (also called hyperbolic numbers), where Lorentz boosts act multiplicatively in a way analogous to ordinary complex rotations~\cite{ca}. Although this provides a suggestive geometric picture, the present paper relies only on standard covering-space arguments, so we do not pursue that formalism here.} The finite coverings are cyclic helicoidal gluings, while the universal cover is the familiar infinite helicoid discussed in~\cite{es}.

\begin{proposition}\label{prop:covertypes}
The universal cover $S_\infty$ is diffeomorphic to $\R^2$. Each finite covering $S_n$ is diffeomorphic to a cylinder.
\end{proposition}

\begin{proof}
Since $\cX$ deformation retracts onto $S^1$, its universal cover is diffeomorphic to the universal cover of a cylinder and therefore to $\R^2$. The finite coverings correspond to finite-index subgroups of $\pi_1(\cX)\cong \Z$, so each $S_n$ has fundamental group isomorphic to $\Z$; a connected two-manifold with that fundamental group is a cylinder.
\end{proof}

\subsection{Lifting the boost action}

The next point is that the boost action on $\cX$ lifts canonically to each connected covering.

\begin{proposition}\label{prop:lifts}
For each $n\in \mathbb{N}\cup\{\infty\}$ the boost $b_\lambda$ lifts to a diffeomorphism $\widehat b_n$ of $S_n$ satisfying
\[
 p_n\circ \widehat b_n=b_\lambda\circ p_n.
\]
For finite $n$, the lift is unique once a sheet containing a chosen lift of a base point in $\Ireg$ is prescribed. The same uniqueness statement holds for $n=\infty$.
\end{proposition}

\begin{proof}
The map $b_\lambda\colon \cX\to\cX$ is homotopic to the identity through the one-parameter family of boosts $b_s$ for $0\le s\le \lambda$. Hence $(b_\lambda)_*$ acts trivially on $\pi_1(\cX)\cong \Z$ and preserves every subgroup $n\Z$. By the covering-space lifting criterion, $b_\lambda$ lifts to each $S_n$. Uniqueness follows from the standard uniqueness of lifts after fixing one point in the fiber over a base point.
\end{proof}

Since $\widehat b_n$ commutes with the deck transformation generating the cyclic covering, the quotient
\begin{equation}\label{eq:Enquot}
 E_n:=S_n/\angles{\widehat b_n},
\end{equation}
with the convention that $E_1=\cX/G$ and $E_\infty=S_\infty/\angles{\widehat b_\infty}$, is well defined. Each $E_n$ is flat away from the images of the chronology horizons and inherits the same local Lorentz metric as Misner spacetime.

%Construction of a helicoid universal cover starting with a stack of punctured Minkowski spaces orbifolded by a Lorentz boost, each corresponding to a copy of a unwrapped maximally extended non-Hausdorff Misner space.

\section{From coverings to genuine extensions}\label{sec:extensions}

The key point is the following: the spaces $S_n$ are merely covering spaces of $\cX$, whereas the quotient spacetimes $E_n$ become genuine extensions of Misner spacetime only after the embedding is exhibited.

Let $j\colon \Ireg\into \cX$ denote the inclusion. Because $\Ireg$ is simply connected, $j$ lifts uniquely to each covering once a base point is fixed.

\begin{lemma}\label{lem:liftI}
For each $n\in\mathbb{N}\cup\{\infty\}$ there exists a unique lift
\[
\widetilde j_n\colon \Ireg\to S_n
\]
of $j$ whose image lies in a distinguished sheet $\Ireg_0\subset S_n$. The map $\widetilde j_n$ is an open isometric embedding.
\end{lemma}

\begin{proof}
Existence and uniqueness follow from the lifting property for coverings and the simple connectedness of $\Ireg$. Since $p_n$ is a local diffeomorphism and a local isometry, the lift of the inclusion is an open isometric embedding.
\end{proof}

Because $\widetilde j_n$ intertwines the actions of $G$ and $\angles{\widehat b_n}$, it descends to the quotient.

\begin{theorem}\label{thm:embedding}
For every $n\in\mathbb{N}\cup\{\infty\}$ there is a well-defined isometric embedding
\[
\iota_n\colon \cM=\Ireg/G\into E_n=S_n/\angles{\widehat b_n}
\]
onto the open subset $\pi_n(\Ireg_0)$, where $\pi_n\colon S_n\to E_n$ is the quotient map. Hence each $E_n$ is a genuine extension of Misner spacetime.
\end{theorem}

\begin{proof}
For $p\in \Ireg$ define
\[
\iota_n([p]_G):=[\widetilde j_n(p)]_{\angles{\widehat b_n}}.
\]
If $p'$ represents the same point of $\cM$, then $p'=b_\lambda^k(p)$ for some $k\in\Z$. Since the lift intertwines the actions, $\widetilde j_n(p')=\widehat b_n^k(\widetilde j_n(p))$, so the definition is independent of the representative. The induced map is smooth, injective, and isometric because it descends from the open isometric embedding $\widetilde j_n$. Its image is open since quotient maps are open and $\widetilde j_n(\Ireg)$ is open in $S_n$.
\end{proof}

Within the class of covering-compatible constructions, this exhausts all possibilities. We emphasize that the following classification result concerns only those extensions obtained by the specific procedure of lifting the fixed boost action to a connected covering of the punctured plane and then taking the quotient.

\begin{theorem}[classification within the covering-compatible class]\label{thm:classification}
Let $p:S\to X$ be a connected covering of the punctured Minkowski plane $X=M^{1,1}\setminus\{Q\}$. Assume that the boost $b_\lambda$ lifts to a diffeomorphism $\hat b$ of $S$, and that the quotient
\[
E:=S/\langle \hat b\rangle
\]
is taken with the descended Lorentz metric. Then, within this covering-compatible construction, there exists a unique $n\in\mathbb N\cup\{\infty\}$ such that $S$ is equivalent, as a covering of $X$, to $S_n$, and $E$ is isometric to $E_n$ by an isometry carrying the embedded copy of Misner spacetime in $E$ onto the embedded copy of Misner spacetime in $E_n$.
\end{theorem}

\begin{proof}
By the classification of connected coverings of $X$, the covering $p$ is equivalent to exactly one of the standard coverings
\[
p_n:S_n\to X,\qquad n\in\mathbb N\cup\{\infty\}.
\]
The hypotheses ensure that we remain inside the class of quotients obtained by lifting the fixed boost action on $X$ and then passing to the quotient by the lifted cyclic action. By \Cref{prop:lifts}, the lifted boost is unique after fixing the sheet containing a chosen lift of a base point in $\Ireg$. Therefore the quotient built from $S$ is identified with the quotient built from $S_n$, and the descended embeddings of $\cM$ agree under this identification.
\end{proof}

\begin{remark}
For $n=1$ one recovers the Hawking--Ellis extension. For $n=\infty$ one obtains the universal-cover extension, whose underlying covering space is simply connected. The intermediate finite values of $n$ give cyclic coverings of the Hawking--Ellis model endowed with the descended boost quotient.
\end{remark}

The phrase ``maximal analytic extension'' should be used carefully \cite{mkg}. The missing origin is a fixed point of the boost action, so adjoining it leads out of the category of smooth manifold quotients.

\begin{proposition}\label{prop:maximality}
Each $E_n$ is maximal within the class of analytic \emph{manifold} extensions obtained by taking a connected covering of $\cX$, lifting the boost action, and quotienting by the lifted action. Any further attempt to adjoin the image of $Q$ destroys the manifold structure.
\end{proposition}

\begin{proof}
The coverings $S_n$ already contain all analytic continuations of the null coordinates away from the missing fixed point. Any enlargement in the same class would have to add a preimage of $Q$. But $Q$ is fixed by the boost, so its image in the quotient has nontrivial isotropy and cannot belong to a smooth manifold quotient. Thus the construction is maximal in the indicated category.
\end{proof}

\section{Causal structure of the family $E_n$}\label{sec:causal}

Each covering $S_n$ inherits lifts of the four wedges of $\cX$; we denote them by
\[
\Ireg_k,\quad \IIreg_k,\quad \IIIreg_k,\quad \IVreg_k,
\]
where $k\in\Z$ for $S_\infty$ and $k\in \Z/n\Z$ for $S_n$. The chronology horizons lift to null boundaries between these sectors.

\begin{figure}
\centering
\includegraphics[width=0.99\columnwidth]{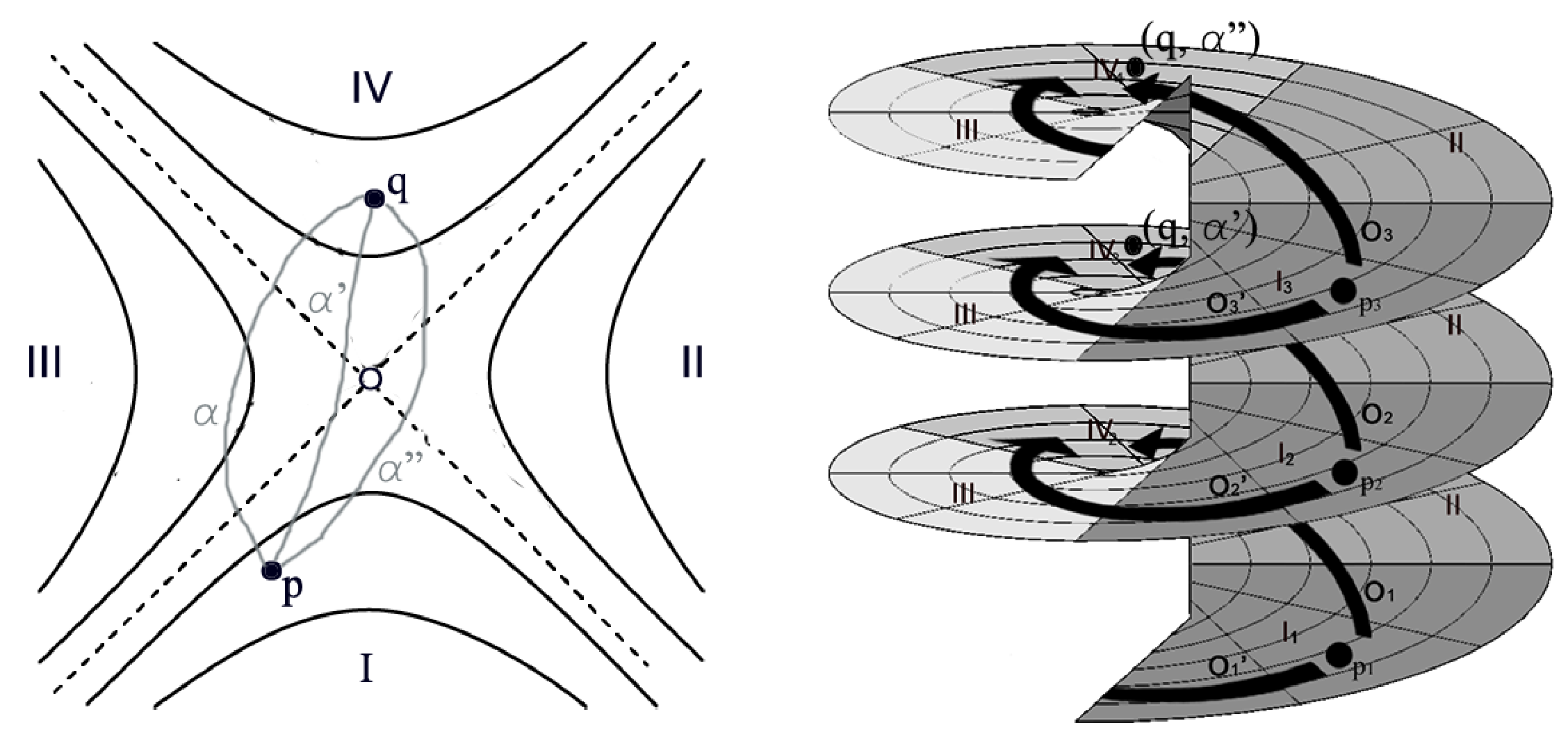}
\caption{Left: two curves from $p$ to $q$ in the punctured plane need not be homotopic, depending on which side of the missing point they pass. All noncontractible
curves in the spacetime on the left are unwrapped to contractible curves in the universal cover. Right: in the universal-cover construction, two observers, $O_{n}$ and $O'_{n}$, that start their journey at the same event $p_{n}$ in region $I_n$ of sheet $n$ won't meet anymore once they have passed the singularity $Q$ on opposite sides as they land in different future chronal sectors. This is the geometric origin of the bi-infinite chain in the universal-cover extension and the cyclic pattern in the finite quotients.}
\label{fig:homotopyhelicoid}
\end{figure}

\medskip
A future-directed timelike curve starting in $\Ireg_k$ can avoid the excised point either by passing through the right dischronal sector $\IIreg_k$ or through the left dischronal sector $\IIIreg_k$. In the universal cover these two possibilities terminate in different future chronal sectors, see \Cref{fig:homotopyhelicoid}. To state this precisely, define the \emph{chronal adjacency digraph} $\Gamma(E_n)$ as follows. Its vertices are the chronal sectors of $E_n$, and there is a directed edge from a chronal sector $C$ to a chronal sector $C'$ if there exists a future-directed timelike curve from some point of $C$ to some point of $C'$.

\begin{theorem}\label{thm:adjacency}
The chronal adjacency digraph of $E_\infty$ is a bi-infinite directed chain, while for finite $n$ the chronal adjacency digraph of $E_n$ is a directed cycle of length $n$.
More precisely:
\begin{enumerate}[label=(\alph*)]
    \item In $E_\infty$, for every $p\in \Ireg_k$ there are future-directed timelike curves from $p$ to points in exactly two future chronal sectors, namely $\IVreg_{k-1}$ and $\IVreg_k$.
    \item In $E_n$ with $n<\infty$, for every $p\in \Ireg_k$ there are future-directed timelike curves from $p$ to points in exactly two future chronal sectors, namely $\IVreg_{k-1}$ and $\IVreg_k$, with indices interpreted modulo $n$.
\end{enumerate}
\end{theorem}

\begin{proof}
Choose the standard cut in the construction of $S_\infty$ along one future null half-line. A future timelike curve from $\Ireg_k$ passing to the right of the missing point does not cross the cut and lands in $\IVreg_k$; one passing to the left crosses the cut once and lands in $\IVreg_{k-1}$. A future-directed timelike curve cannot cross a chronology horizon and then return to the original lifted Minkowski chart while remaining timelike; hence no other future chronal sector is reachable from the same initial chronal sector. This proves part~(a).

\medskip
Part~(b) is obtained by quotienting the sheet index modulo $n$. The same two local timelike routes around the missing point remain, but the bi-infinite indexing becomes periodic. Therefore the adjacency digraph is the cyclic quotient of the chain, namely a directed cycle of length $n$.
\end{proof}

\begin{definition}
Two spacetimes $E$ and $E'$ will be called \emph{causally isomorphic} if there exists a bijection
\[
F:E\to E'
\]
such that, for all points $p,q\in E$,
\[
p\ll q \quad\Longleftrightarrow\quad F(p)\ll F(q),
\]
where $\ll$ denotes the chronological relation.
\end{definition}

\begin{corollary}\label{cor:notcausalisomorphic}
If $m\neq n$ in $\mathbb{N}\cup\{\infty\}$, then $E_m$ and $E_n$ are not causally isomorphic.
\end{corollary}

\begin{proof}
Any causal isomorphism preserves the decomposition into chronal sectors and preserves the future-reachability relation defining the adjacency digraph. By \Cref{thm:adjacency}, $\Gamma(E_m)$ and $\Gamma(E_n)$ are nonisomorphic when $m\neq n$. Therefore $E_m$ and $E_n$ cannot be causally isomorphic.
\end{proof}

The local geometry of each $E_n$ is flat, so geodesics are projections of Minkowski geodesics on the corresponding covering space. The only obstruction to completeness is the missing fixed point.

\begin{proposition}\label{prop:geodesics}
Every inextendible geodesic in $E_n$ is of one of the following two types: either it has infinite affine parameter in both directions, or one of its lifts to $S_n$ reaches, in finite affine parameter, a missing preimage of the excised fixed point $Q$. Equivalently, affine incompleteness in $E_n$ is caused only by the ideal ends corresponding to the omitted preimages of $Q$. Passing from $E_1$ to the higher coverings introduces no new curvature singularity.
\end{proposition}

\begin{proof}
Lift the geodesic to $S_n$. Since $S_n$ is flat away from the omitted preimages of $Q$, the lift is a straight null, timelike, or spacelike geodesic in a Minkowski chart. Hence it extends uniquely until either it has unbounded affine parameter or it reaches, in finite affine parameter, one of the deleted preimages of $Q$. Projecting back to the quotient preserves affine incompleteness and does not create curvature, since the quotient is locally flat away from the chronology horizons and the omitted ideal ends.
\end{proof}

\section{Isocausality with two-dimensional Schwarzschild metrics}\label{sec:isocausality}

We next compare Misner-type spacetimes with two-dimensional Schwarzschild-type metrics.

\subsection{Definitions and setup}

Following García-Parrado and Senovilla~\cite{gps,gs}, a smooth map $F\colon (X,g)\to (Y,h)$ is \emph{causal} if $dF$ sends every future-directed causal vector of $g$ to a future-directed causal vector of $h$. Two spacetimes are \emph{isocausal} if there exist diffeomorphisms in both directions that are causal. The case of Misner space and pseudo-Schwarzschild spacetime has already been settled: their isocausality was proved in~\cite{rieger} (the pseudo-Schwarzschild spacetime was introduced and analyzed in~\cite{ori}).

~

Consider the two-dimensional Schwarzschild metric
\begin{equation}\label{eq:schw}
 g_{\mathrm{Sch}}=-f(r)\,dt^2+f(r)^{-1}\,dr^2,
 \qquad
 f(r)=1-\frac{2m}{r},
 \qquad r>0.
\end{equation}
On the exterior region $r>2m$ and the interior region $0<r<2m$, introducing the Regge--Wheeler coordinate $r_*$ via $dr_*=f(r)^{-1}dr$ gives
\[
 g_{\mathrm{Sch}}=f(r)\,(-dt^2+dr_*^2).
\]
Thus every simply connected region avoiding $r=2m$ is conformal to a Minkowski domain.

\subsection{Local comparison}

The chronal region of Misner spacetime is locally a flat Lorentzian cylinder, so local isocausality is immediate.

\begin{proposition}\label{prop:localiso}
Every point of Misner spacetime and every point of each extension $E_n$ admits a neighborhood that is isocausal to a neighborhood of a point of the two-dimensional Schwarzschild exterior. The same holds for neighborhoods away from the singular endpoint in the Schwarzschild interior.
\end{proposition}

\begin{proof}
Both metrics are two-dimensional Lorentz metrics and therefore locally conformally flat. On a sufficiently small simply connected neighborhood there are null coordinates in which each metric is a positive conformal multiple of $-du\,dv$. In dimension two a positive conformal rescaling preserves the null cones exactly, so the identity map between the corresponding Minkowski-coordinate domains is causal in both directions.
\end{proof}

\subsection{A global obstruction}

The dischronal regions of Misner-type extensions contain closed timelike curves, whereas the standard Schwarzschild exterior and interior are chronological. This gives an immediate obstruction.

\begin{theorem}\label{thm:noiso}
No extension $E_n$ containing a dischronal region is globally isocausal to a chronological region of two-dimensional Schwarzschild spacetime. In particular, no such $E_n$ is globally isocausal to the Schwarzschild exterior $\{r>2m\}$ or to the interior $\{0<r<2m\}$ with its standard time orientation.
\end{theorem}

\begin{proof}
Suppose $F\colon E_n\to Y$ were a causal diffeomorphism onto a chronological Schwarzschild region $Y$. Let $\gamma$ be a future-directed timelike closed curve in the dischronal part of $E_n$. Then $F\circ \gamma$ is a future-directed causal closed curve in $Y$, contradicting chronology. Therefore no such causal diffeomorphism exists, and hence no isocausality exists.
\end{proof}

\subsection{The chronal Misner region}

The original Misner cylinder $\cM=\Ireg/G$ is chronological but not simply connected. Since $\Ireg$ is conformal to a Minkowski half-plane and the Schwarzschild exterior is conformal to a strip in $(t,r_*)$-coordinates, one can still compare compact subregions.

\begin{proposition}\label{prop:compactiso}
For every relatively compact subregion $K\subset \cM$ there exists a relatively compact region $K'\subset \{r>2m\}$ in the two-dimensional Schwarzschild exterior such that $K$ and $K'$ are isocausal.
\end{proposition}

\begin{proof}
Lift $K$ to a relatively compact subset $\widetilde K\subset \Ireg$. Since the quotient map $\Ireg\to\cM$ is locally an isometry, $\widetilde K$ is isocausal to $K$. In null coordinates on a compact Schwarzschild rectangle with conformal factor bounded above and below away from zero, the identity map between the corresponding Minkowski rectangle and the Schwarzschild rectangle is causal in both directions. Choosing the rectangle large enough to contain an isometric copy of $\widetilde K$ gives the claim.
\end{proof}

\section{Discussion and conclusion}

The punctured Minkowski plane $X=M^{1,1}\setminus\{Q\}$ is not itself an extension of Misner spacetime. Rather, it is the common covering model from which the extensions arise only after one lifts the boost action to a connected covering of $X$, forms the corresponding quotient, and then exhibits the embedding of  \Cref{thm:embedding}. Within that natural covering-compatible class, \Cref{thm:classification} shows that the resulting family is exactly $\{E_n\}_{n\in\mathbb{N}\cup\{\infty\}}$. The family is not distinguished merely by fundamental group. \Cref{thm:adjacency} shows that the causal arrangement of chronal sectors changes from an $n$-cycle to a bi-infinite chain; this produces a concrete causal invariant and yields the nonisomorphism statement of \Cref{cor:notcausalisomorphic}. The geodesic behavior remains governed entirely by the missing fixed point, and no new curvature singularity appears in the higher coverings.

\medskip
The isocausality analysis clarifies both the usefulness and the limitations of causal comparison. On chronal simply connected regions, Misner and Schwarzschild-type metrics are locally indistinguishable from the causal viewpoint because every two-dimensional Lorentz metric is locally conformally flat. Globally, however, chronology violation is decisive: no Misner-type extension with closed timelike curves can be globally isocausal to a chronological Schwarzschild region. The already established Misner--pseudo-Schwarzschild isocausality~\cite{rieger} fits naturally into this picture.

\medskip
Several questions remain open. One may ask for a classification of non-Hausdorff flat extensions of Misner spacetime that do not arise from coverings of the punctured plane, or for analogous constructions in higher-dimensional Misner-type quotients and in pseudo-Schwarzschild models. The present work provides a clean two-dimensional template for such questions.

\section*{Acknowledgments}

I am greatly indebted to Kip S.\ Thorne for his supportive remarks, encouragement and enthusiasm. His guidance and suggestions about the structure and content of this paper were invaluable during this research.

\section*{Declarations}

\noindent \text{Conflict of Interest:} The authors declare that they have no conflicts of interest.

\smallskip

\noindent \text{Ethical Statement:} This article is a purely theoretical work; no ethical approval was required for the research described.

\smallskip

\noindent \text{Data Availability:} Data sharing is not applicable to this article as no datasets were generated or analyzed during the current study.

\end{document}